\documentclass[aps,prl,reprint,superscriptaddress,nofootinbib]{revtex4-1}
\usepackage{graphicx}
\usepackage{dcolumn}
\usepackage{bm}
\usepackage{braket}
\usepackage{epstopdf}
\usepackage{amsmath}
\usepackage{hyperref}
\usepackage{color}
\usepackage{cancel}

\begin{document}
\raisebox{20pt}[0pt][0pt]{\hspace*{20mm} RIKEN-iTHEMS-Report-23, YITP-23-14}

\title{Doubly Charmed Tetraquark $T^+_{cc}$ from Lattice QCD near Physical Point }

\author{Yan Lyu}
\email{helvetia@pku.edu.cn}
\affiliation{State Key Laboratory of Nuclear Physics and Technology, School of Physics, Peking University, Beijing 100871, China }
\affiliation{Interdisciplinary Theoretical and Mathematical Sciences Program (iTHEMS), RIKEN, Wako 351-0198, Japan}
\author{Sinya Aoki}
\email{saoki@yukawa.kyoto-u.ac.jp}
\affiliation{Center for Gravitational Physics, Yukawa Institute for Theoretical Physics, Kyoto University, Kyoto 606-8502, Japan}
\affiliation{Interdisciplinary Theoretical and Mathematical Sciences Program (iTHEMS), RIKEN, Wako 351-0198, Japan}
\author{Takumi Doi}
\email{doi@ribf.riken.jp}
\affiliation{Interdisciplinary Theoretical and Mathematical Sciences Program (iTHEMS), RIKEN, Wako 351-0198, Japan}
\author{Tetsuo Hatsuda}
\email{thatsuda@riken.jp}
\affiliation{Interdisciplinary Theoretical and Mathematical Sciences Program (iTHEMS), RIKEN, Wako 351-0198, Japan}
\author{Yoichi Ikeda}
\email{yikeda@cider.osaka-u.ac.jp}
\affiliation{Center for Infectious Disease Education and Research, Osaka University, Suita 565-0871, Japan}
\author{Jie Meng}
\email{mengj@pku.edu.cn}
\affiliation{State Key Laboratory of Nuclear Physics and Technology, School of Physics, Peking University, Beijing 100871, China }
\affiliation{Yukawa Institute for Theoretical Physics, Kyoto University, Kyoto 606-8502, Japan}

\date{\today}
\begin{abstract}
The doubly charmed tetraquark $T^+_{cc}$ recently discovered by the LHCb Collaboration is studied on the basis of $(2+1)$-flavor lattice QCD simulations of the $D^*D$ system with nearly physical pion mass $m_\pi=146$ MeV.
The interaction of $D^*D$ in the isoscalar and $S$-wave channel, derived from the hadronic spacetime correlation by the HAL QCD method, is attractive for all distances and leads to a near-threshold virtual state with a pole position $E_\text{pole}=-59\left(^{+53}_{-99}\right)\left(^{+2}_{-67}\right)$ keV and  a large scattering length $1/a_0=0.05(5)\left(^{+2}_{-2}\right)~\text{fm}^{-1}$.
The virtual state is shown to evolve into a loosely bound state as $m_\pi$ decreases to its physical value by using a potential modified to $m_\pi=135$ MeV based on the pion-exchange interaction. Such a potential
is found to give a semiquantitative  description of the  LHCb data on the $D^0D^0\pi^+$ mass spectrum.
Future study is necessary to perform physical-point simulations with the isospin-breaking and open three-body-channel effects taken into account.
\end{abstract}


\maketitle

{\it Introduction.$-$}
The quest of exotic hadrons with multiquark configuration beyond the conventional constituent quark model 
 has been one of central subjects in the study of nonperturbative QCD for decades~\cite{Chen2016,Esposito2016,Lebed2016,Ali2017,Guo2017,Yamaguchi2019}.
Although dozens of candidates of exotic hadrons were reported~\cite{Olsen2017, Brambilla2019}, a doubly charmed tetraquark  $T^+_{cc}$ was discovered only recently by the LHCb Collaboration~\cite{LHCb2021_NP}: A pronounced narrow peak appears in the $D^0D^0\pi^+$ mass spectrum just around $360$ keV below the $D^{*+}D^0$ threshold, and its isospin $I$, spin $J$, and
 parity $P$ are found to be consistent with $(I,J^P)=(0,1^+)$~\cite{LHCb2021}.

Although early theoretical predictions on the mass of $T^+_{cc}$ were scattered in  the range of $\pm$300 MeV with respect to the $D^{*+}D^0$ threshold, the constraint from the LHCb data
 starts to lead a consistent description on the basic properties of $T^+_{cc}$ in recent 
  phenomenological models (see Refs.~\cite{Ader1981,Lipkin1986,Zouzou1986,Janc2004,Ebert2007,Esposito:2013fma,Karliner2017,Du2021,Ling2021,Kim2022,Agaev:2022ast,Chen2022_PPP}, and references therein).	
 Nevertheless, a solid theoretical validation on the existence of $T^+_{cc}$ needs to be 
 given from first-principles lattice QCD simulations. 
 So far, the lattice QCD studies on the doubly charmed tetraquark 
 in terms of the $D^{*}D$ scattering length~\cite{Ikeda2013, Chen2022, Padmanath2022}
 and the finite-volume $D^{*}D$ energy ~\cite{Cheung2017,Junnarkar2018} 
have been limited in the region of  large pion masses ($m_\pi\geq280$ MeV)
 and small lattice sizes ($L\leq 2.9$ fm).

\begin{figure}[htbp]
  \centering
  \includegraphics[width=8cm]{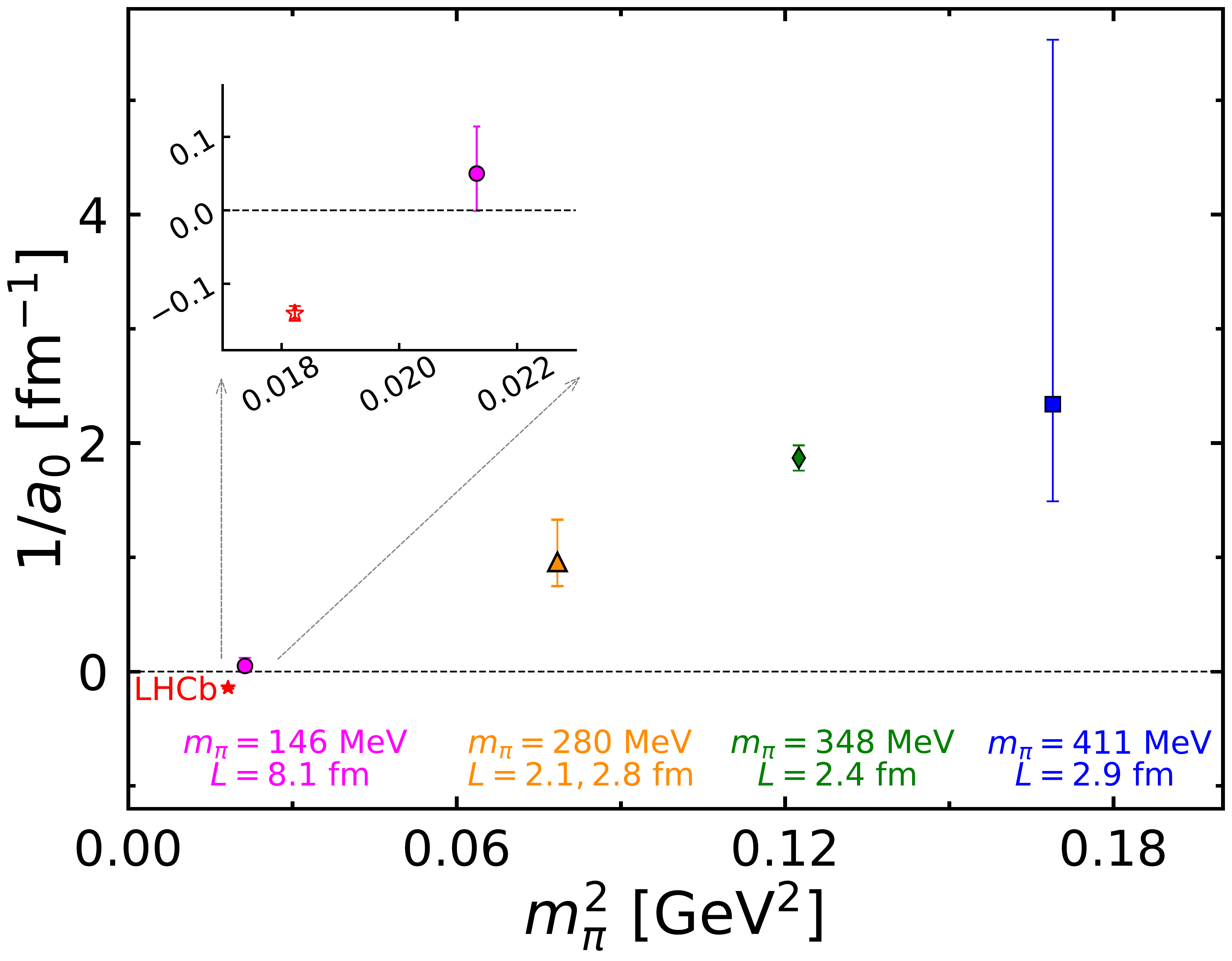}
  \caption{The inverse of scattering length $1/a_0$ for the $D^*D$ scattering in the $I=0$ and $S$-wave channel obtained from lattice QCD simulations by Refs.~\cite{Ikeda2013}(blue square),~\cite{Chen2022}(green diamond), and ~\cite{Padmanath2022}(yellow triangle).
 The result of the present Letter (magenta circle) and the real part of the experimental value by LHCb  (red star)~\cite{LHCb2021} are also shown.
  } \label{Fig-a0}
\end{figure}

Under the above circumstances, the purpose of this Letter is twofold.
 The first purpose is to report an investigation of the $D^*D$ scattering in the $I=0$ and $S$-wave channel from lattice QCD simulations with nearly physical pion mass $m_\pi=146$ MeV and a large lattice size $L=8.1$ fm.  We employ the HAL QCD method~\cite{Ishii2007, Ishii2012, Aoki2020} for converting the hadronic spacetime correlation to the physical observables.
 Shown in Fig.~\ref{Fig-a0} is a summary of the previous lattice QCD calculations of the inverse scattering length of $D^{*}D$ as a function of $m_{\pi}^2$.
 The recent LHCb result and the result of the present Letter are also shown.  
  Figure~\ref{Fig-a0} indicates a clear trend that the lattice data approach the experimental data as the pion mass decreases toward the physical point.
 Our second purpose is 
 to make a direct  comparison between theoretical and  experimental  $D^0D^0\pi^+$ mass spectra 
 to unveil the properties of near-threshold tetraquark. 
 The HAL QCD method is best suited for such comprehensive analysis, since it provides the $D^* D$ scattering $T$ matrix and  
allows us to study how the mass spectrum changes toward the physical point.

{\it HAL QCD method.$-$}
The starting point for deriving $D^*D$ interaction in the HAL QCD method is the normalized spacetime correlation ~\cite{Ishii2007, Ishii2012, Aoki2020},
\begin{eqnarray}\label{Eq-R}
R(\bm r,t)&=&\sum\limits_{\bm{x}}\braket{0|D^*(\bm x+\bm r,t)D(\bm x,t)\overline{\mathcal{J}}(0)|0}/e^{-(m_{D^*}+m_{D})t}  \nonumber \\
&=&\sum\limits_n A_n \psi_n(\bm r)e^{-(\Delta E_n)t} + O(e^{-(\Delta E^*)t}),
\end{eqnarray}
where  $\Delta E_n=\sqrt{m_{D^*}+\bm{k}_n^2}+\sqrt{m_{D}+\bm{k}_n^2} - (m_{D^*}+m_{D})$ with
$\bm{k}_n$ being relative momentum in the center-of-mass frame.
The equal-time Nambu-Bethe-Salpeter wave function  for each elastic scattering state is defined as $\psi_n(\bm r)$,
and the coupling strength to the $n$th eigenstate is denoted by $A_n=\braket{E_n|\overline{\mathcal{J}}(0)|0}$. 
We use local sink operators of the form $\bar{q}(x)\Gamma c(x)$ with $\Gamma=\gamma_i \ (\gamma_5) $ for $D^* \ (D)$, where $q(x)$ denotes either $u(x)$ or $d(x)$.
For the source, the wall-type operator $\mathcal{J}$ at $t=0$, defined by replacing $q(\bm x, t)$ with $\sum_{\bm x}q(\bm x, t)$ and $c(\bm x, t)$ with $\sum_{\bm x}c(\bm x, t)$, is adopted together with the Coulomb gauge fixing.
Although the wall-type source may have weak coupling to compact states,  it is a suitable operator for studying near-threshold and loosely bound $T^+_{cc}$.
By projecting into quark zero momentum mode, such a source operator is expected to have large overlap with the low-energy $D^*D$ scattering states. 
The lowest relevant (noninteracting) energy levels of three-body inelastic state ($DD\pi$) and two-body inelastic state ($D^*D^*$)
 are  $\Delta E^*\simeq \sqrt{m_D^2+(2\pi/L)^2}+\sqrt{m_\pi^2+(2\pi/L)^2}-m_{D^*}=78$ MeV and
 $\Delta E^*\simeq m_{D^*}-m_D=140$ MeV, respectively. They are suppressed by  the combination of the 
  small overlap with our wall-type source operator and the factor $e^{-(\Delta E^*)t}$.
The $S$-wave projection of $R(\bm r,t)$ on the lattice
is carried out by the Misner method~\cite{Miyamoto2020}. 
A comparison between $S$- and $D$- wave components of $R(\bm r,t)$ is given in Fig. S1 in~\cite{Supp},
and
we find the $D$-wave component is highly suppressed to only $\lesssim~O(0.1)\%$, similar to the observations in Refs.~\cite{Padmanath2022,Du2021}.

The correlation function $R(\bm r,t)$ is known to satisfy the partial differential equation ~\cite{Ishii2012,Aoki:2011gt},
\begin{eqnarray}\label{Eq-U}
 \left[ \frac{1+3\delta^2}{8\mu}{\partial_t^2}-{\partial_t}-H_0+O(\delta^2\partial^3_t)\right]&R(\bm r,t)&  \ \ \ \ \nonumber \\
=\int d\bm{r}'U(\bm r,\bm r')&R(\bm r',t),&
\end{eqnarray}
where $H_0=\frac{-\nabla^2}{2\mu}$, $\mu=\frac{m_{D^*}m_{D}}{m_{D^*}+m_{D}}$, $\delta=\frac{m_{D^*}-m_{D}}{m_{D^*}+m_{D}}$. The integral kernel $U(\bm r,\bm r')$ is defined through $[\bm{k}_n^2/(2\mu) -H_0]\psi_n(\bm r) =\int d\bm {r}'U(\bm r, \bm r’) \psi_n(\bm r')$.
The $O(\delta^2\partial^3_t)$ term is found to be consistent with zero within statistical error and is neglected  in our analysis.
The derivative expansion $U(\bm r,\bm r')=\sum_nV_n(\bm r)\nabla^n\delta(\bm r-\bm r')$ at the leading order gives an effective local potential,
\begin{eqnarray}\label{Eq-V}
V(r)= R^{-1}(\bm r,t) \left[ \frac{1+3\delta^2}{8\mu}{\partial_t^2}-{\partial_t}-H_0\right]R(\bm r,t).
\end{eqnarray}
The truncation error from higher-order terms of the derivative expansion 
 can be estimated through the $t$ dependence of $V(r)$~\cite{Ishii2012, Iritani2019Jhep, Lyu2022}.
Once the potential is obtained, it can be used to calculate physical quantities 
by solving the stationary Schr\"odinger equation, $[H_0+V(r)]\psi (\bm r) = E\psi (\bm r)$, in the infinite spatial volume.

{\it Lattice setup.$-$}
The $(2+1)$-flavor gauge configurations are generated on the $96^4$ lattice with the Iwasaki gauge action and the nonperturbatively $O(a)$-improved Wilson quark action at nearly physical point ($m_\pi=146.4$ MeV)~\cite{Ishikawa2016}. The lattice spacing is $a=0.0846$ fm, corresponding to lattice size $L=8.1$ fm.
For the charm quark, the relativistic heavy quark (RHQ) action is employed in order to remove the cutoff errors
associated with the charm quark mass up to next-to-next-to-leading order~\cite{Aoki2003}.
The charm quark mass is set to be very close to its physical value, which leads to a spin-averaged $1S$ charmonium mass $M_\text{av}\equiv(m_{\eta_c} +3m_{J/\Psi})/4=3096.6$ MeV, $0.9\%$ larger than the physical value~\cite{Namekawa2017}.
By comparing results from another set of RHQ parameters corresponding to $M_\text{av}=3051.4$ MeV, $0.6\%$ smaller than the physical value, we confirm that effect from slightly unphysical charm quark is small compared with current statistical uncertainties.

With 200 gauge configurations, we perform 640 measurements ($=$ 4 directions $\times$ 80 source positions $\times$ 2 forward-backward propagations)~\cite{Lyu2022_Nphi} for each configuration to increase the statistics.
The jackknife method and bootstrap method are used to estimate the statistical error with a bin size of 20 configurations throughout this Letter; a comparison with a bin size of 40  shows that the bin size dependence is small.
The quark propagators are calculated by the domain-decomposed solver~\cite{Ishikawa2023} and the Bridge++ code~\cite{Bridge} with the periodic boundary condition for all directions. The unified contraction algorithm is used to obtain the hadronic correlation functions~\cite{Doi2013}.
The $D^*$ and $D$ masses from the correlated single-state fit in the temporal region $t/a=20-30$ are $m_{D^*}=2018.1(5)$ MeV and $m_D=1878.2(2)$ MeV with statistical error shown in the parenthesis, which are $0.5\%$ heavier than the physical values. (For the effective mass plots, see  Fig. S2 in \cite{Supp}.)
 With our slightly heavy pion mass,  $m_{D^*}$ is below the $D\pi$ threshold, so that
$D^*$ is stable against the strong decay.  Other systematic errors can be estimated as follows: (i) The finite cutoff effect is  $O[(a\Lambda_\text{QCD})^2, \alpha^2_sa\Lambda_\text{QCD}]\simeq O(1)\%$ thanks to $O(a)$ improvement for the light quarks and RHQ action for the charm quark,
(ii) the finite volume effect is as small as $\exp[-m_h(L/2)]\simeq0.3\%$ (where $m_h=2m_\pi$ as described below) thanks to the large volume, and (iii) the quenched charm quark effect is expected to be highly suppressed by the heavy charm quark mass~\cite{Cali:2019enm}.

\begin{figure}[t]
  \centering
  \includegraphics[width=8cm]{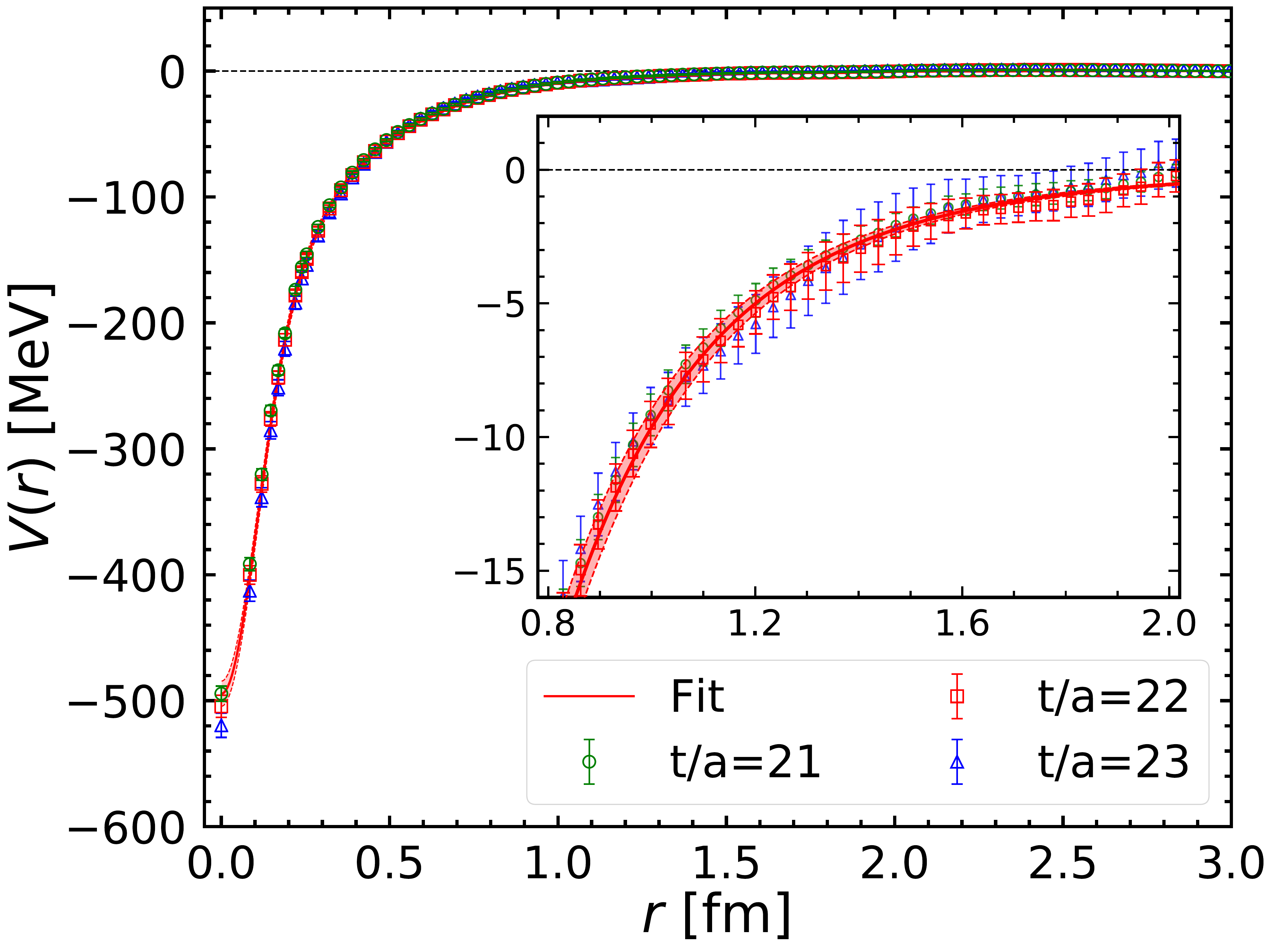}
  \caption{The $D^*D$ potential $V(r)$ in the $I=0$ and  $S$-wave channel at Euclidean time $t/a= 21$ (green circles), $22$ (red squares), and $23$ (blue triangles). The red band shows the fitted potential with $V^B_{\rm fit}$ for $t/a=22$. The inset shows a magnification.
  } \label{Fig-V}
\end{figure}

{\it Interaction potential.$-$}
We show in Fig.~\ref{Fig-V} the $D^*D$ potential $V(r)$ in the $I=0$ and $S$-wave channel defined in Eq.~(\ref{Eq-V}) for $t/a=21$, $22$, and $23$, corresponding to $t\simeq1.9$ fm. 
This temporal region is chosen to suppress inelastic states contamination at smaller $t$ and simultaneously to avoid large statistical errors at larger $t$.
A small variation of the potentials for different $t/a$ is observed and is taken into account as a source of the systematic error.
 
For later convenience, we perform a correlated fit to the potential in Fig.~\ref{Fig-V} in the range $0<r<2$~fm by a phenomenological four-range Gaussian, 
$V^A_\text{fit}(r)=\sum^4\limits_{i=1}a_ie^{-(r/b_i)^2}$. 
Fitting parameters at  $t/a=22$ are $(a_1, a_2, a_3, a_4)=[-269(6), -121(10), -81(12), -23(14)]$ in MeV and  $(b_1, b_2, b_3, b_4)=[0.14(1), 0.27(1), 0.52(5), 0.97(16)]$ in fm with an accuracy of $\chi^2/\text{d.o.f.}=1.01$.
The fitted potential is shown in Fig. S3 in \cite{Supp} with the
normalized covariance matrix of the fitted parameters given in Eq.(S2) in \cite{Supp}.

The $D^* D$ potential in the $(I,J^P)=(0,1^+)$ channel shown in Fig.~\ref{Fig-V}  is attractive for all distances.
 This is consistent with the result previously found for heavy pion masses ~\cite{Ikeda2013}: The short-range attraction was suggested to be related to the attractive (anti)diquark configuration
$(\bar{u}\bar{d})_{{\bm 3}_c, I=J=0}-(cc)_{{\bm 3}^*_c, J=1}$,
   coupled to the asymptotic $D^*D$ state \cite{Jaffe:2003sg, Selem2006,Lee:2009rt, Deng2022, Noh2021}~(for an interpretation based on the string model, see Ref.~\cite{Andreev:2021eyj}). 
   Similar short-range attraction was also found for
   the  $B^*B$ system in the  $(I,J^P)=(0,1^+)$ channel~\cite{Aoki2022,Bicudo2015}. 
The long-range part of the attraction for $r > 1$ fm would 
have contributions from the one-pion exchange (OPE) between $D^*$ and $D$ of the form $\sim {e^{-m r}}/{r}$  with either $m=m_\pi$~\cite{Ohkoda2012} or $m=\sqrt{(m_{D^*}-m_D)^2-m^2_\pi}$~\cite{Li2012}, and from  the two-pion exchange (TPE) of the form $\sim  ({e^{-m_{\pi} r}}/r)^2$ \cite{Lyu2022_Nphi}.  

 To study  different possibilities for the long-range part, we introduce the following fit function 
  with the Gaussian-type form factor \cite{Lyu2022_Nphi}:
\begin{eqnarray}
V^B_\text{fit}(r;m_{\pi})=\sum\limits_{i=1,2}&a_i&e^{- \left({r}/{b_i}\right)^2}\\ \nonumber
 +&a_3&\left(1-e^{- \left({r}/{b_3}\right)^2}\right)^n V_\pi^n(r)
\end{eqnarray}
 with $V_\pi(r)={e^{-m_{\pi} r}}/{r}$.
 We find that $n=2$ and $m_\pi=146.4$ MeV provide a best fit with the parameter set, $(a_1, a_2)=[-276(6), -219(8)]$ in MeV, $a_3= -43(3)~\text{MeV}\cdot\text{fm}^2$, and  $(b_1, b_2, b_3)=[0.14(1), 0.28(1), 0.43(2)]$ in fm, with an accuracy of $\chi^2/\text{d.o.f.}=0.96$.
 The fitted potential is shown by the red band in Fig.~\ref{Fig-V}.
  Also, we find that neither $n=1$ and $m_\pi=146.4$ MeV nor $n=1$ and $m_{\pi} \rightarrow \sqrt{(m_{D^*}-m_D)^2-m^2_\pi}$
  can reproduce the long-range part of the potential.
  In Fig.~S4 in~\cite{Supp},
   the spatial effective energy $E_\text{eff}(r)=-\frac{\ln [V(r)r^2/a_3]}{r}$ with 
    the lattice data for $V(r)$ and $a_3$ as inputs is shown to have a plateau at $2m_\pi=292.8$ MeV for $1 < r  < 2$ fm. 
    This indicates that the long-range part is consistent with the TPE.
  It is an open question why the theoretically possible OPE contribution does not appear in the lattice data.

\begin{figure}[t]
  \centering
  \includegraphics[width=8cm]{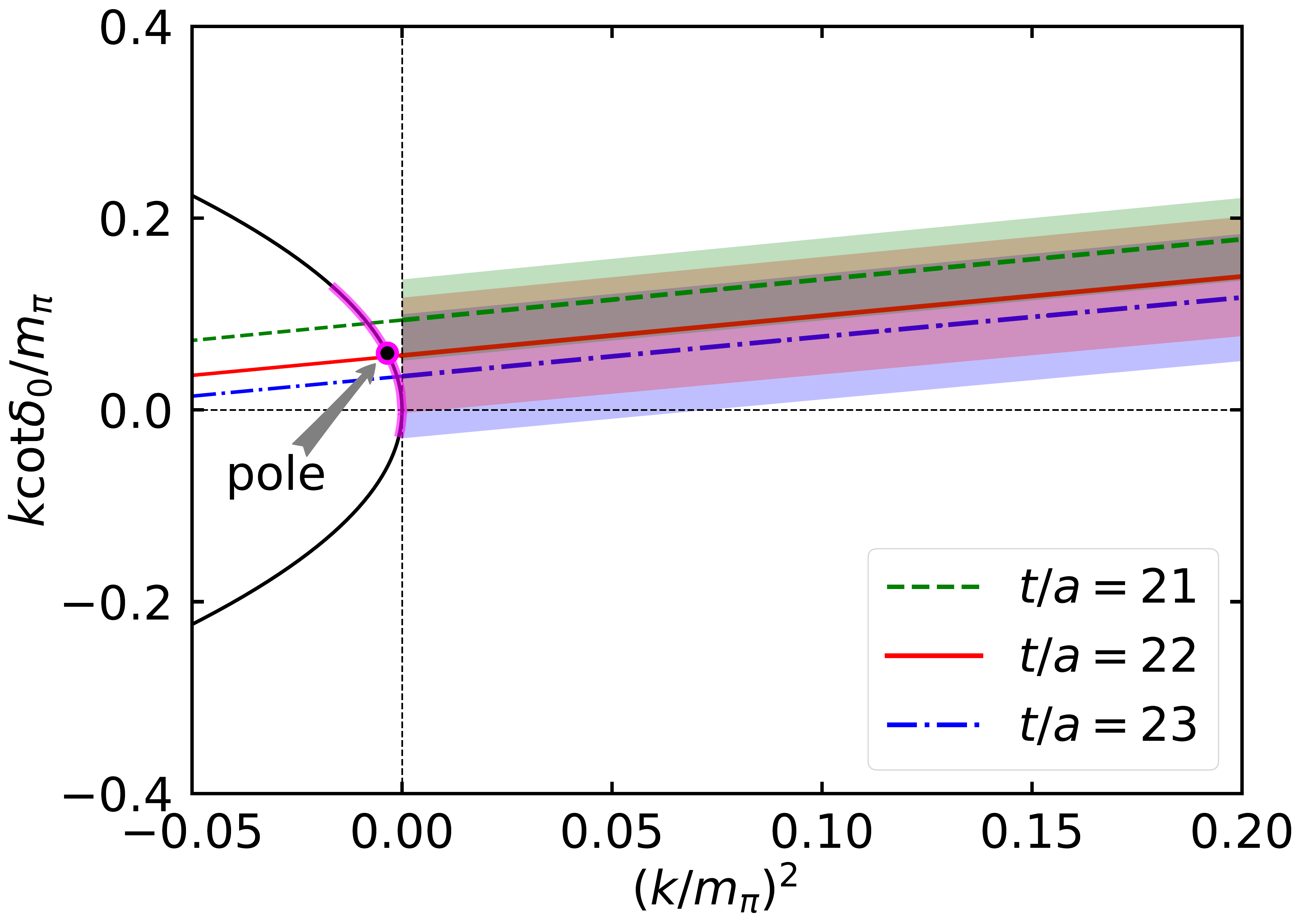}
  \caption{The $k\cot\delta_0/m_\pi$ for $D^*D$ scattering in the $I=0$ and $S$-wave channel as a function of $(k/m_\pi)^2$.  The intersection of $k\cot\delta_0/m_\pi$ and the black solid line $[\pm\sqrt{-(k/m_\pi)^2}]$
  denotes the pole of the scattering amplitude. The pole position is shown by the black point with magenta line indicating statistical and systematic errors combined.
  } \label{Fig-ERE}
\end{figure}
\begin{table}[b]
\caption{Results for  $1/a_0$, the effective range $r_{\rm eff}$, the pole position $\kappa_\text{pole}$, and $E_\text{pole}$.
Numbers in the second column with statistical error (first parentheses) and systematic error (second parentheses) are obtained from $V^{A,B}_{\rm fit}(r)$ with $t/a=21-23$ at $m_\pi=146.4$ MeV.
 The third column shows estimated values from $V^B_{\rm fit}(r;m_{\pi})$ with $t/a=22$ and $m_\pi=135.0$ MeV.
The asymmetric statistical error for $E_{\rm pole}$ is due to its non-normal distribution  (see Fig. S5 in ~\cite{Supp}).}
\begin{tabular}{lcc}
  \hline\hline
    $m_\pi$ (MeV)~~~~~~~~~&$146.4$~~~~~~~~~&$135.0$\\
  \hline
$1/a_0$ (fm$^{-1}$) ~~~~~~~~~&$0.05(5)\left(^{+2}_{-2}\right)$ ~~~~~~~~~~&$-0.03(4)$ \\
$r_{\rm eff}$ (fm) ~~~~~~~~~&$1.12(3)\left(^{+3}_{-8}\right)$ ~~~~~~~~~~&$1.12(3)$ \\
$\kappa_\text{pole}$ (MeV) ~~~~~~~~~&$-8(8)\left(^{+3}_{-5}\right)$ ~~~~~~~~~~&$+5(8)$ \\
$E_\text{pole}$ (keV) ~~~~~~~~~&$-59\left(^{+53}_{-99}\right)\left(^{+2}_{-67}\right)$~~~~~~~~~~&$-45\left(^{+41}_{-78}\right)$\\
 \hline\hline
\end{tabular} \label{tab-scattering}
\end{table}

{\it Scattering parameters and pole position.$-$}
Using the potential fitted to our lattice data, we calculate the $S$-wave scattering phase shifts $\delta_0$ by solving the Schr\"odinger equation in the infinite spatial volume with $m_{D^{(*)}}$ measured on the lattice. Figure~\ref{Fig-ERE} shows the $k\cot\delta_0/m_\pi$ as a function of $(k/m_\pi)^2$ with $k$ being a relative momentum.
The scattering length $a_0$ and the effective range $r_\text{eff}$ are obtained by an effective-range expansion (with the sign convention of high-energy physics) as
\begin{eqnarray}
    k\cot\delta_0=\frac{1}{a_0} +\frac12r_\text{eff}k^2+O(k^4),
\end{eqnarray}
and are given in the second column of 
 Table~\ref{tab-scattering}.
The central values and the statistical errors in the first parentheses are obtained at $t/a=22$ with $V^B_\text{fit}$, while the systematic errors in the second parentheses are obtained by comparing results from different $t/a=21-23$ with $V^{A,B}_\text{fit}$.

The scattering length obtained in this way is shown by the magenta circle in Fig.~\ref{Fig-a0} together with the previous lattice 
results and the LHCb experimental data.
 As mentioned in the {\it Introduction}, there is a clear tendency that $1/a_0$ from lattice data approaches the unitary regime ($1/a_0 \sim 0)$ as the pion mass decreases.
 Also, our result at $m_{\pi}=146.4$ MeV produces a virtual state, which
  implies that a marginal modification of the interaction (e.g. by reducing the quark mass) may bring 
 such a near-threshold virtual state to a loosely bound state.
A typical example of virtual or bound state sensitive to the quark mass is
dineutron or deuteron in nuclear physics~\cite{Inoue2012, Horz2021, Amarasinghe2021}.
A bound (virtual) state is characterized by a pole of 
the scattering amplitude $f(k)$ on the positive (negative) side of the imaginary axis 
at $k=i\kappa_\text{pole}$.
Since  $f^{-1}(k)$ is proportional to $k \cot \delta_0$,  
 virtual and bound  poles near threshold can be inferred from the intersection between $\frac{1}{a_0} +\frac12r_\text{eff}k^2$ and $\pm\sqrt{-k^2}$. As shown in Fig.~\ref{Fig-ERE},  we indeed find a near-threshold virtual state pole.
The actual value of the pole position in the complex $k$ plane 
for the present pion mass is shown in the second column of Table~\ref{tab-scattering}, 
together with the pole energy, 
$E_\text{pole} =\sqrt{m^2_{D^*}-\kappa^2_\text{pole}}+\sqrt{m^2_{D}-\kappa^2_\text{pole}}-(m_{D^*}+m_D)$.
We confirm that $\kappa_\text{pole}$ and $E_\text{pole}$ here have numerically the same values  with those
 directly obtained from the $T$ matrix, indicating that the left-hand cut associated with TPE is negligible.
The effect on $E_\text{pole}$ from slightly unphysical charm quark mass is found to be about $-30$ keV, smaller than statistical and systematic errors in Table~\ref{tab-scattering}.

To estimate how the scattering parameters change and the pole evolves toward the physical point,
 we modify the potential 
 by taking $m_{\pi}=135.0$ MeV (approximately the physical pion mass without the QED contribution~\cite{FLAG:2021npn}) 
  with the other parameters ($a_{1,2,3}, b_{1,2,3}$)  of $V^B_\text{fit}$ at $t/a=22$ kept fixed.
Using such a potential together with physical $m_{D^{*+}, D^0}$~\cite{PDG2020}
we find a loosely bound state  with the scattering parameters and pole positions 
 given in the third column of Table~\ref{tab-scattering}. This indicates the existence of a bound $T^+_{cc}$ at physical point,
 though there is still a  quantitative difference from the experimental value $E_\text{pole}=-360(40)\left(^{+4}_{-0}\right)$ keV reported by LHCb Collaboration~\cite{LHCb2021}.  
Since the above estimate does not account for the isospin-breaking effect nor the open three-body-channel effect, 
future works should directly perform $(1+1+1)$-flavor or $(1+1+1+1)$-flavor lattice QCD $+$ QED simulations with physical quark masses and the 
 three-body channels ($D^0D^0\pi^+$ and $D^0D^+\pi^0$) considered.

Let us make an alternative estimate of $a_0$ at the physical point by taking
the present and previous lattice data shown in Fig.~\ref{Fig-a0} (note that these data are from different calculations and possess different lattice systematics):
By using the simplest fit function $1/a_0(m_\pi)=c+d m_\pi^2$,
 we find $1/a_0=-0.01(9)~\text{fm}^{-1}$  for $m_\pi=135.0$ MeV ~(Fig. S6 in \cite{Supp}).
This result is consistent with $1/a_0= -0.03(4)$ fm$^{-1}$ obtained by $V^{B}_{\rm fit}$ with the same pion mass
in Table I, supporting the validity of our modification procedure for $V^{B}_{\rm fit}$.

\begin{figure}[htbp]
  \centering
  \includegraphics[width=8.7cm]{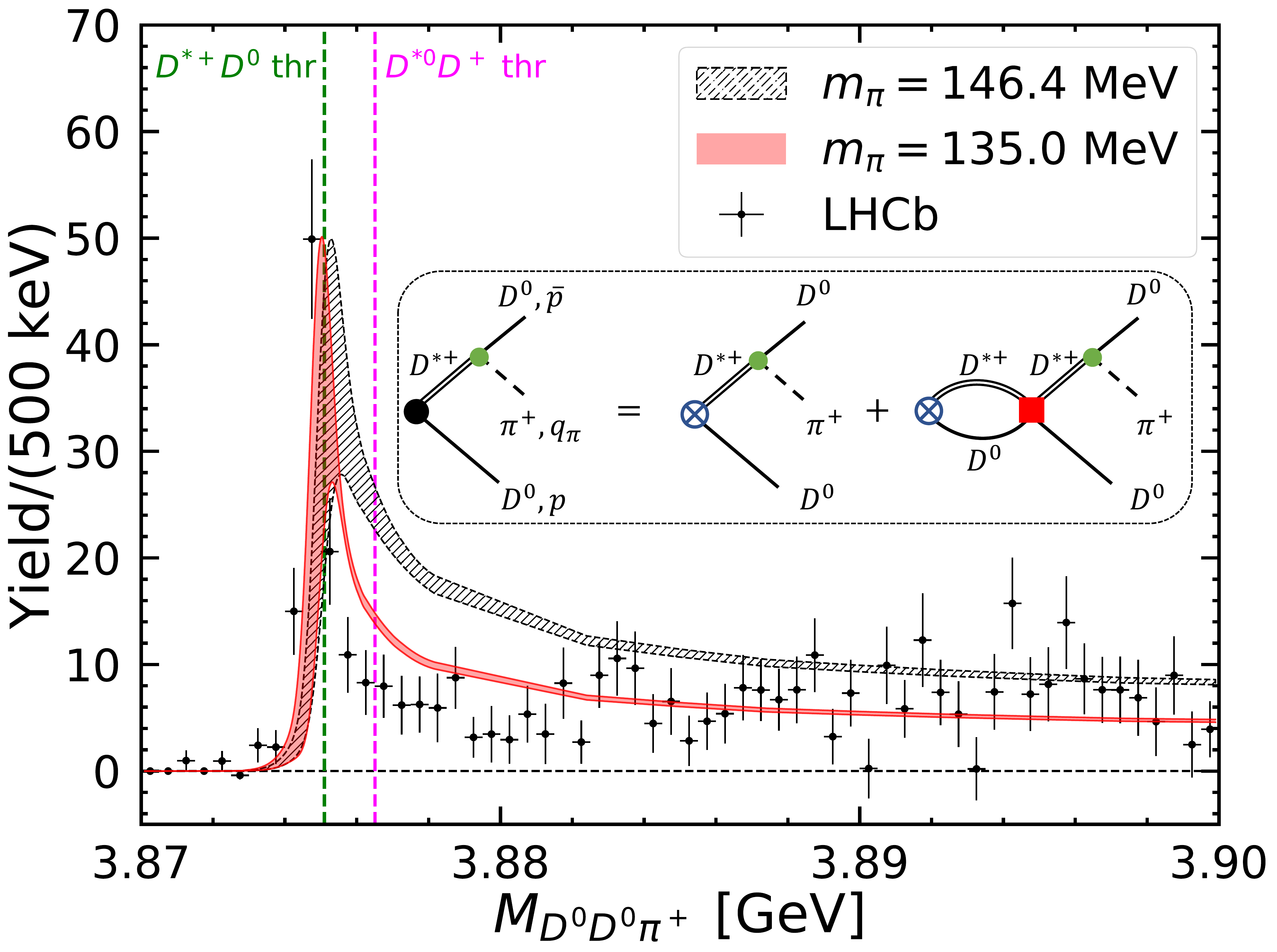}
  \caption{The $D^0D^0\pi^+$ mass spectrum. Theoretical results with $V^B_\text{fit}(r; m_\pi)$ for $m_\pi=146.4$ MeV ($m_\pi=135.0$ MeV) are shown by the black (red) band.
The black points are LHCb data~\cite{LHCb2021}. 
The inset shows diagrams contributing to the $D^0D^0\pi^+$ mass spectrum, where the black filled circle, blue cross circle, green filled circle, and red square denote production amplitude $U$, constant vertex $P$, $D^{*+}\rightarrow D^0\pi^+$ vertex, and scattering $T$ matrix, respectively.
  } \label{Fig-LS}
\end{figure}

{\it $D^0D^0\pi^+$ mass spectrum.$-$}
In order to make a further connection to the LHCb experimental data, let us now construct $D^0D^0\pi^+$ mass spectrum based on  the above interaction by considering the rescattering between $D^{*+}$ and $D^0$ along the line with Refs.~\cite{Du2021,Albaladejo2021}.
Within a single channel framework, we first construct a production amplitude $U(M, p)$ for $D^{*+}D^0$ pair with invariant mass $M$ and relative momentum $p$ in the $I=0$ and $S$-wave channel generated from a constant vertex $P$. Then the amplitude consisting of a direct production process and a rescattering process is written as (see the inset in Fig.~\ref{Fig-LS}),
\begin{eqnarray}
 U(M,p)=P + \int\frac{d^3\bm{q}}{(2\pi)^3}T(M,p,q)G(M,q)P.
\end{eqnarray} 
Here the $T$ matrix $T(M,p,q)$ with incoming (outgoing) momentum $q$ $(p)$ is obtained by solving the Lippmann-Schwinger equation with a given potential. 
 The $D^{*+}D^0$ propagator is denoted as $G(M,q)=\left(M-m_{D^{*+}}-m_{D^0}-\frac{q^2}{2\mu}+\frac{i}{2}\Gamma_{D^{*+}}\right)^{-1}$  where $\Gamma_{D^{*+}}=82.5$ keV is the decay width of $D^{*+}$~\cite{PDG2020}, representing the unstable nature of $D^{*+}$ in the real world.
We set $P=1$ without loss of generality, since it can be absorbed into an overall normalization factor.
Then, the $D^0D^0\pi^+$ mass spectrum reads
\begin{eqnarray}\label{Eq-M}
 & & \frac{d\text{Br}[D^0D^0\pi^+]}{dM}=\mathcal{N}\int pdp \int \bar{p}d\bar{p} \\ \nonumber
 & & \ \ \ \ \ \ |U(M,p)G(M,p)q_\pi(p) + U(M,\bar{p})G(M,\bar{p})q_\pi(\bar{p}) |^2 ,
\end{eqnarray} 
where the pion momentum $q_\pi(p)$ arises from $D^{*+}\rightarrow D^0\pi^+$ vertex, and can be determined kinematically~\cite{Du2021}.
The second term of the integrand with $\bar{p}$ comes from symmetrizing the $D^{0} D^{0} \pi^+$ amplitude due to two identical $D^{0}$s in the final state.
With $\mathcal{N}$ being an overall normalization factor, the shape of $D^0D^0\pi^+$ mass spectrum
does not depend on any free parameters.
In order to compare with the LHCb data, an energy resolution function given in Ref.~\cite{LHCb2021} is convolved with Eq.~(\ref{Eq-M}).

Shown in Fig.~\ref{Fig-LS} by the black and red bands are our theoretical calculations for $D^0D^0\pi^+$ mass spectrum obtained from $V^B_\text{fit}(r; m_\pi)$ with $m_\pi=146.4$ MeV and $m_\pi=135.0$ MeV, respectively.
 In both cases,   experimental values for $m_{D^{*+}, D^0, \pi^+}$ ~\cite{PDG2020} are adopted
 in the kinematics in order to keep the same phase space with the experiment.
Shown together by the black points are LHCb data~\cite{LHCb2021}.
The obtained mass spectrum has a pronounced peak around the $D^{*+}D^0$ threshold,
 and the peak position shifts to the left as $m_\pi$ decreases; this is consistent with the  evolution from a near-threshold virtual pole to a loosely bound pole. 
At the physical pion mass, the red band with a peak just on the $D^{*+}D^0$ threshold 
provides a better description of the LHCb data, though visible differences still exist.  
This calls for a direct calculation of the potential with the physical quark masses as well as with the isospin-breaking effect.

{\it Summary and discussion.$-$}
In this Letter, we present an investigation on the scattering properties of the $D^*D$ system based on the $(2+1)$-flavor lattice QCD simulations with nearly physical pion mass $m_\pi=146$ MeV.
The attractive potential between $D^*$ and $D$ in the $I=0$ and $S$-wave channel is extracted from hadronic spacetime correlation. The long-range part of the potential is found to be dominated by the two-pion exchange
 at least in the range $1 < r < 2 $ fm. The overall attraction 
 is found to be strong enough to lead a near-threshold virtual state with a pole position $E_\text{pole}=-59\left(^{+53}_{-99}\right)\left(^{+2}_{-67}\right)$ keV and  a large scattering length $1/a_0=0.05(5)\left(^{+2}_{-2}\right)~\text{fm}^{-1}$.
The virtual state is shown to evolve into a loosely bound state at the physical point by using a potential modified to $m_\pi=135$ MeV based on the TPE.
Such a potential can also provide a semiquantitative description to the $D^0D^0\pi^+$ mass spectrum measured by LHCb Collaboration.

We are currently under way to perform physical-point simulations in (2+1)-flavor QCD.
It is also important in the future to study the OPE and the associated left-hand cut as well as  to
perform high-precision study with the isospin-breaking effect and open three-body-channel effect by $(1+1+1)$-flavor or $(1+1+1+1)$-flavor lattice QCD $+$ QED calculations with the physical quark masses.

\begin{acknowledgments}
{\it Acknowledgments.$-$}
We thank members of the HAL QCD Collaboration for stimulating discussions.
Y.L. thanks Meng-Lin Du, and Xu Feng for the helpful discussions.  We thank members of the PACS Collaboration for the gauge configuration generation conducted on the K computer at RIKEN.
The lattice QCD measurements have been performed on Fugaku and HOKUSAI supercomputers at RIKEN.
We thank ILDG/JLDG \cite{ldg}, which serves as essential infrastructure in this study.
This work was partially supported by HPCI System Research Project (hp120281, hp130023, hp140209, hp150223, hp150262, hp160211, hp170230, hp170170, hp180117, hp190103, hp200130, hp210165, hp210212, hp220240, hp220066, and hp220174),  the National Key R\&D Program of China (Contracts No. 2017YFE0116700 and No. 2018YFA0404400), the National Natural Science Foundation of China (Grants No. 11935003, No. 11975031, No. 11875075, No. 12070131001, and No. 12141501), the JSPS (Grants No. JP18H05236, No. JP22H00129, No. JP19K03879, No. JP18H05407, No. JP21K03555, and No. JP23H05439),  ``Priority Issue on Post-K computer'' (Elucidation of the Fundamental Laws and Evolution of the Universe), ``Program for Promoting Researches on the Supercomputer Fugaku'' (Simulation for basic science: from fundamental laws of particles to creation of nuclei) 
and (Simulation for basic science: approaching the new quantum era), 
and Joint Institute for Computational Fundamental Science (JICFuS).

\end{acknowledgments}


\
\onecolumngrid
\newpage
\renewcommand{\thefigure}{S\arabic{figure}}
\renewcommand{\theequation}{S\arabic{equation}}
\renewcommand{\thetable}{S\arabic{table}}
\setcounter{figure}{0}
\setcounter{equation}{0}
\setcounter{table}{0}

\begin{center}
\fontsize{12pt}{15pt}{\bf{SUPPLEMENTAL MATERIAL}}
\end{center}

\

We present  the $S$- and $D$-wave components of $R(\bm r,t)$, the effective mass of $D^{(*)}$, a detailed analysis of the potential, the distribution of $E_{\rm pole}$, and a chiral extrapolation of the inverse of scattering length ($1/a_0$) in this supplemental material.

\subsection{The $S$- and $D$-wave components of $R(r, t)$}
\begin{figure}[htbp]
  \centering
  \includegraphics[width=8cm]{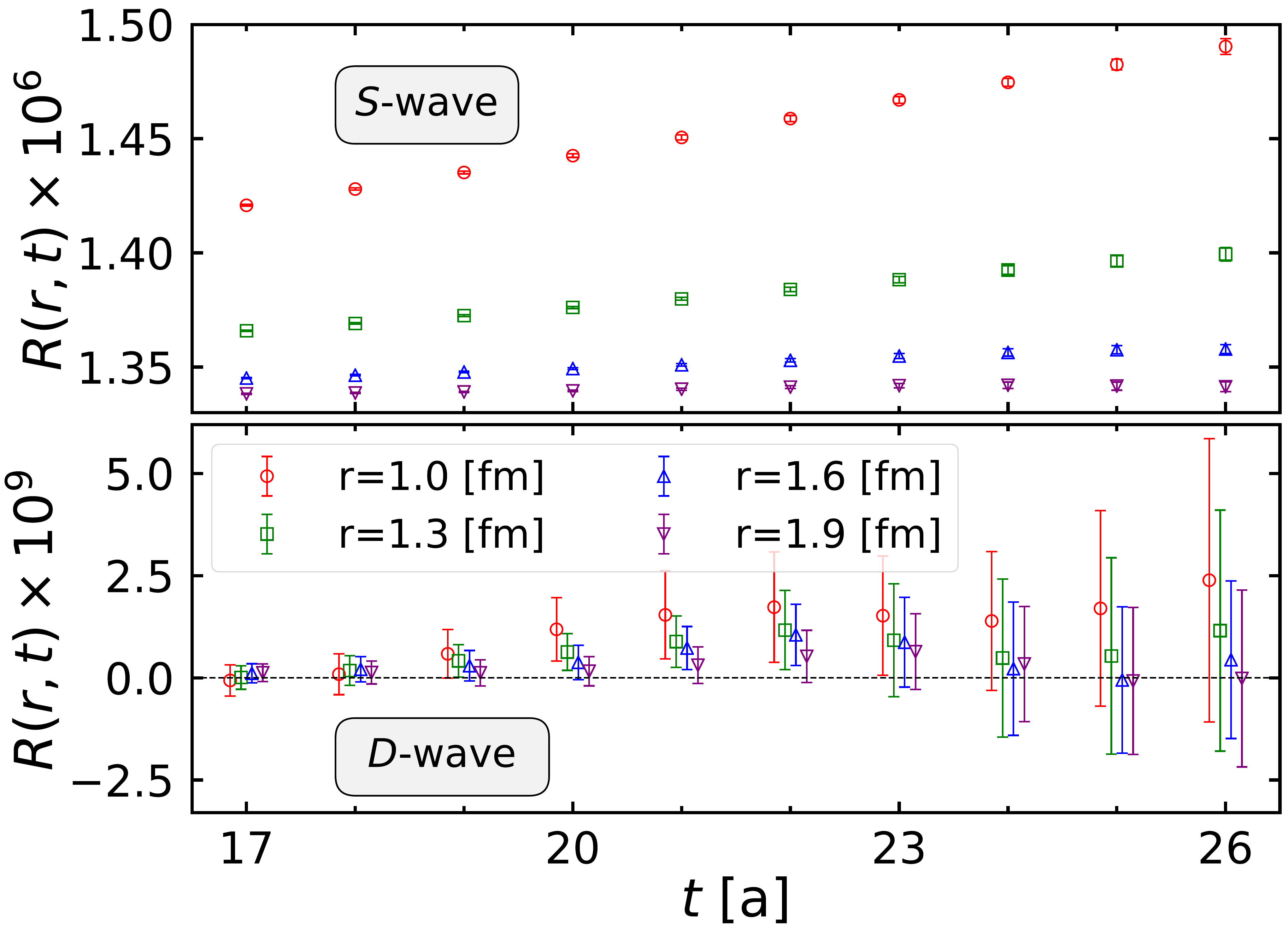}
  \caption{
 The $S$- (upper) and $D$- (lower) wave components of $R(\bm r,t)$ as a function of $t$ for several values of $r$ where two-pion- exchange (TPE) is relevant.
  } \label{Supp-Fig_Rcorr}
\end{figure}

We show the $S$- and $D$-wave components of $R(r, t)$ in Fig. \ref{Supp-Fig_Rcorr}.
The $D$-wave component is around three order-of-magnitude smaller than the $S$-wave component, and is consistent with zero within the statistical error.

\newpage

\subsection{The effective mass}
\begin{figure}[htbp]
  \centering
  \includegraphics[width=8cm]{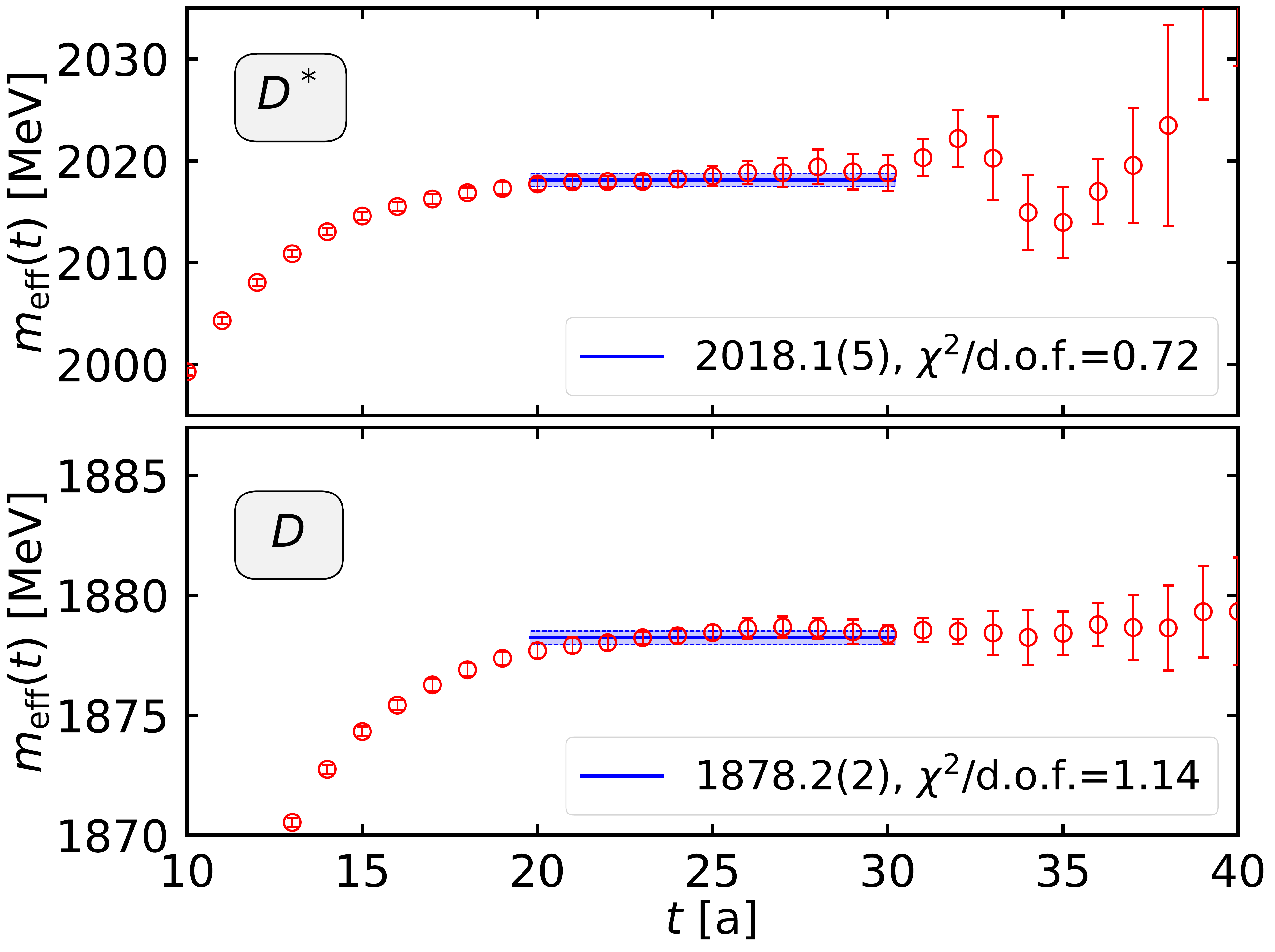}
  \caption{
 The effective mass for $D^*$ (upper) and $D$ (lower).
 The blue bands show the results of correlated single-state fit.
  } \label{Supp-Fig_meff}
\end{figure}

The effective masses for $D^*$ and $D$, defined as $m_{\rm eff}(t)=\ln\frac{C(t)}{C(t+1)}$ with $C(t)$ being the two-piont function, are shown in Fig.~\ref{Supp-Fig_meff}.

\subsection{A detailed analysis of the potential}
\begin{figure}[htbp]
  \centering
  \includegraphics[width=16cm]{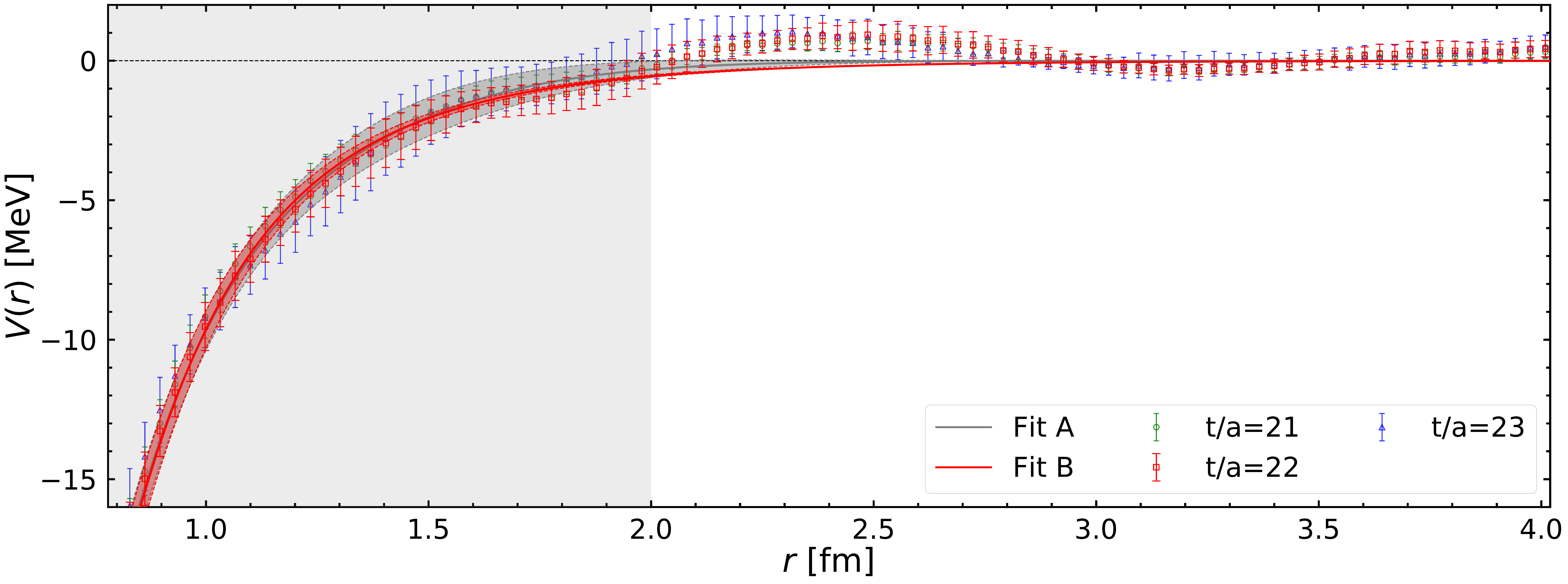}
  \caption{
 The $D^*D$ potential $V(r)$ in the $I=0$ and  $S$-wave channel at Euclidean time $t/a= 21$ (green circles), $22$ (red squares), and $23$ (blue triangles). The gray (red) band shows the fitted potential with $V^A_{\rm fit}$ ($V^B_{\rm fit}$) for $t/a=22$. Fit range is $0<r<2$ fm. 
  } \label{Supp-Fig_V}
\end{figure}

In Fig.~\ref{Supp-Fig_V}, we show the long-range potential up to $r=4$ fm (half of the box size) together with two fits as bands.

\begin{figure}[htbp]
  \centering
  \includegraphics[width=8cm]{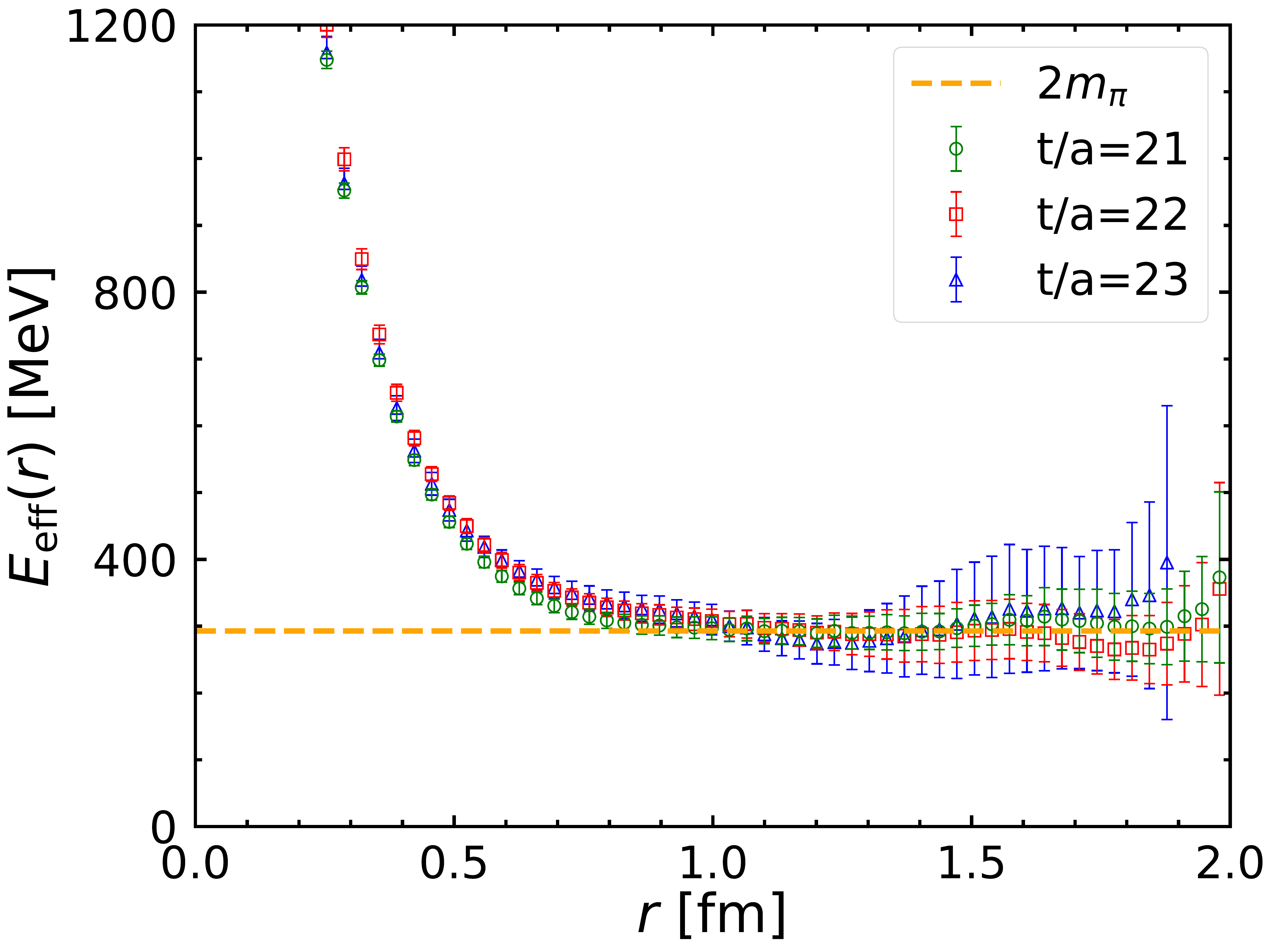}
  \caption{
 The spatial effective energy $E_\text{eff}(r)$ as a function of separation $r$ at Euclidean time $t/a= 21$ (green circles), $22$ (red squares), and $23$ (blue triangles).  
The orange dashed line denotes $2m_\pi$ with $m_\pi=146.4$ MeV.
  } \label{Supp-Fig1}
\end{figure}

We show in Fig.~\ref{Supp-Fig1} the spatial effective energy defined as,
\begin{equation}
E_\text{eff}(r)=-\frac{\ln [V(r)r^2/a_3]}{r},
\end{equation}
which has a plateau at $2m_\pi=292.8$ MeV for $r>1.0$ fm. 
This numerically shows that the long-range potential is consistent with the TPE form $\sim\frac{e^{-2m_\pi r}}{r^2}$.

The normalized covariance matrix, defined as $\frac{{\rm Cov}(x,y)}{\sigma_x \sigma_y}$ with ${\rm Cov}(x,y)$ being the covariance between $x$ and $y$ and $\sigma_{x(y)}$ being the standard error of $x(y)$, of the fitted parameters $(a_1,b_1,\cdots,a_4,b_4)$ in $V^A_{\rm fit}$ are
\begin{eqnarray}
    \left[\begin{array}{rrrrrrrr}
        1.00 & -0.58 & 0.43 & -0.80 & -0.40 & -0.28 &-0.10 &0.07 \\
          & 1.00 & -0.39 & 0.77 & 0.76 & 0.54 & 0.12 & -0.08 \\
          &   &  1.00  & -0.66 & -0.08 & -0.79 & -0.68 & -0.64 \\
          &   &     &  1.00 & 0.48 & 0.68 & 0.43 & 0.23 \\
          &   &     &    &  1.00   & 0.03  & -0.46 & -0.59 \\
          &   &     &    &      &   1.00   & 0.85 & 0.69 \\
          &   &     &    &      &       &   1.00  & 0.93 \\
          &   &     &    &      &       &      &  1.00 \\
    \end{array}\right],
\end{eqnarray}
and $(a_1,b_1,\cdots,a_3,b_3)$ in $V^B_{\rm fit}$ are
\begin{eqnarray}
    \left[\begin{array}{rrrrrr}
        1.00 & -0.49 & -0.03 & -0.63 & 0.09 & -0.18 \\
          & 1.00 & 0.68 & 0.66 & 0.21 & 0.53  \\
          &   &  1.00  & 0.14 & -0.29 & -0.03 \\
          &   &     &  1.00 & 0.36 & 0.81 \\
          &   &     &    &  1.00   & 0.80 \\
          &   &     &    &      &   1.00   \\
    \end{array}\right].
\end{eqnarray}

\begin{figure}[htbp]
  \centering
  \includegraphics[width=7.5cm]{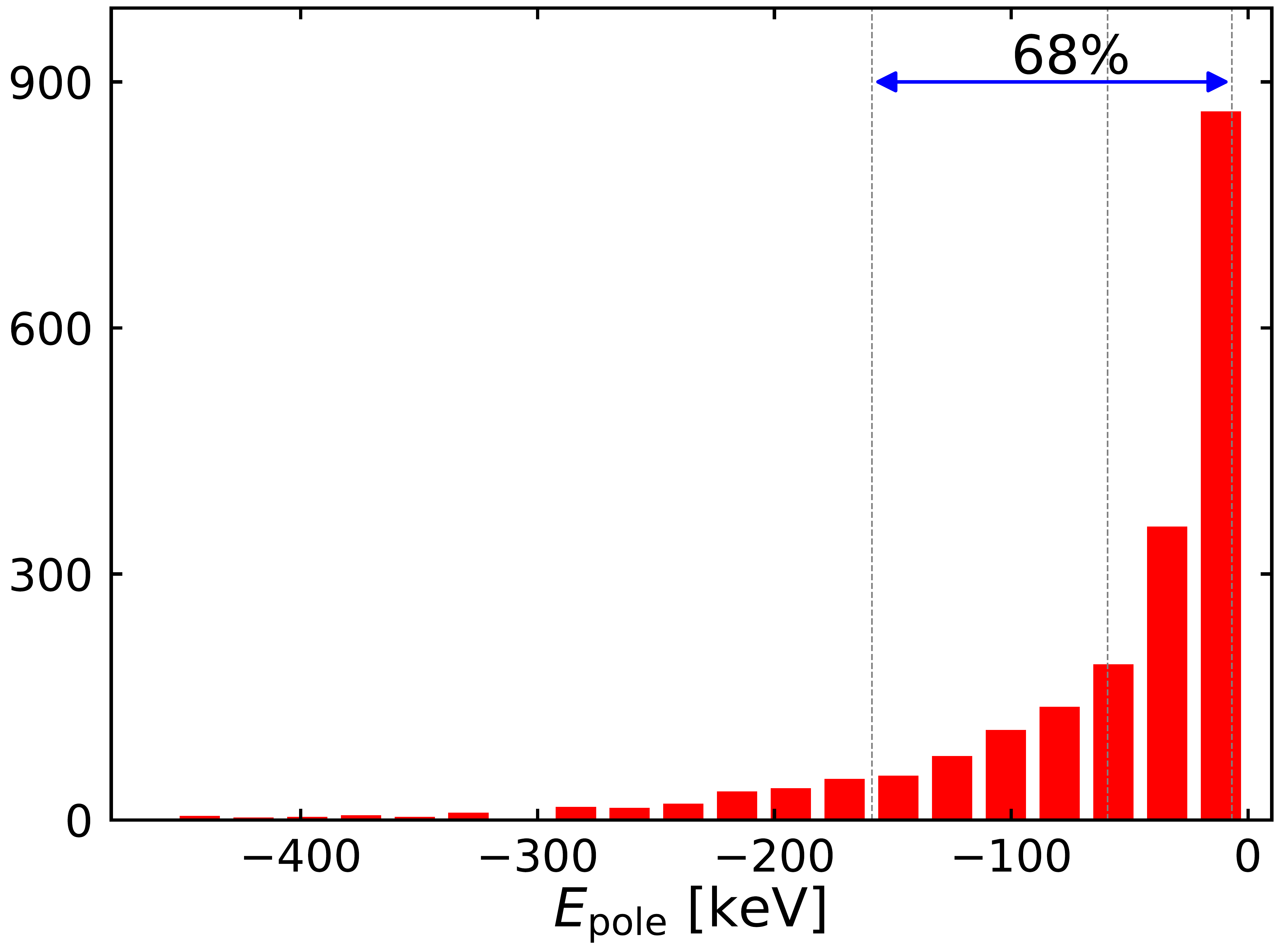}
  \caption{
The histogram of the PDF of $E_{\rm pole}$. $\mu-a$, $\mu$, and $\mu+b$ are also shown as dashed lines.}
\label{Supp-Epole}
\end{figure}

\subsection{The distribution of $E_{\rm pole}$}

$E_{\rm pole}$ is defined as $E_{\rm pole} = \sqrt{m^2_{D^*} -\kappa_{\rm pole}^2}+\sqrt{m^2_{D} -\kappa_{\rm pole}^2}-(m_{D^*} + m_D)$ and is less than or equal to zero.
The probability distribution function (PDF) of $E_{\rm pole}$ from our lattice simulations is shown in Fig.~\ref{Supp-Epole}.
For such an asymmetric distribution, we define the statistical error by demanding $P(\mu-a<E_{\rm pole}<\mu+b)=68\%$, $P(E_{\rm pole}<\mu-a)=32\%\times P(E_{\rm pole}<\mu)$ 
and $P(E_{\rm pole}>\mu+b)=32\%\times P(E_{\rm pole}>\mu)$
with $\mu$ and $a$ ($b$) being the mean value and the lower (upper) bound.
Then, $E_{\rm pole}=\mu\left(^{+b}_{-a}\right)=-59\left(^{+53}_{-99}\right)$ keV is quoted.

\subsection{A chiral extrapolation of $1/a_0$}
\begin{table}[tbhp]
\caption{The fit parameters in Eq.~(\ref{eq-mpi-fit}) with $1\sigma$ uncertainty quoted in the parenthesis. The corresponding value for $1/a_0$ at $m_\pi=135.0$ MeV is shown in the last column.}
\begin{tabular}{cccc}
  \hline\hline
    ~~~~~~$c$ [fm$^{-1}$]~~~~~~&$d$~~~~~~&$\chi^2/\text{d.o.f.}$~~~~~~&$1/a_0$ [fm$^{-1}$]\\
  \hline
 ~~~~~~$-0.33(6)$~~~~~~&$18(1)~\text{GeV}^{-2}\cdot\text{fm}^{-1}$~~~~~~&$0.1$~~~~~~&$-0.01(9)$ \\
 \hline\hline
\end{tabular} \label{tab-mpi-fit}
\end{table}

\begin{figure}[htbp]
  \centering
  \includegraphics[width=8cm]{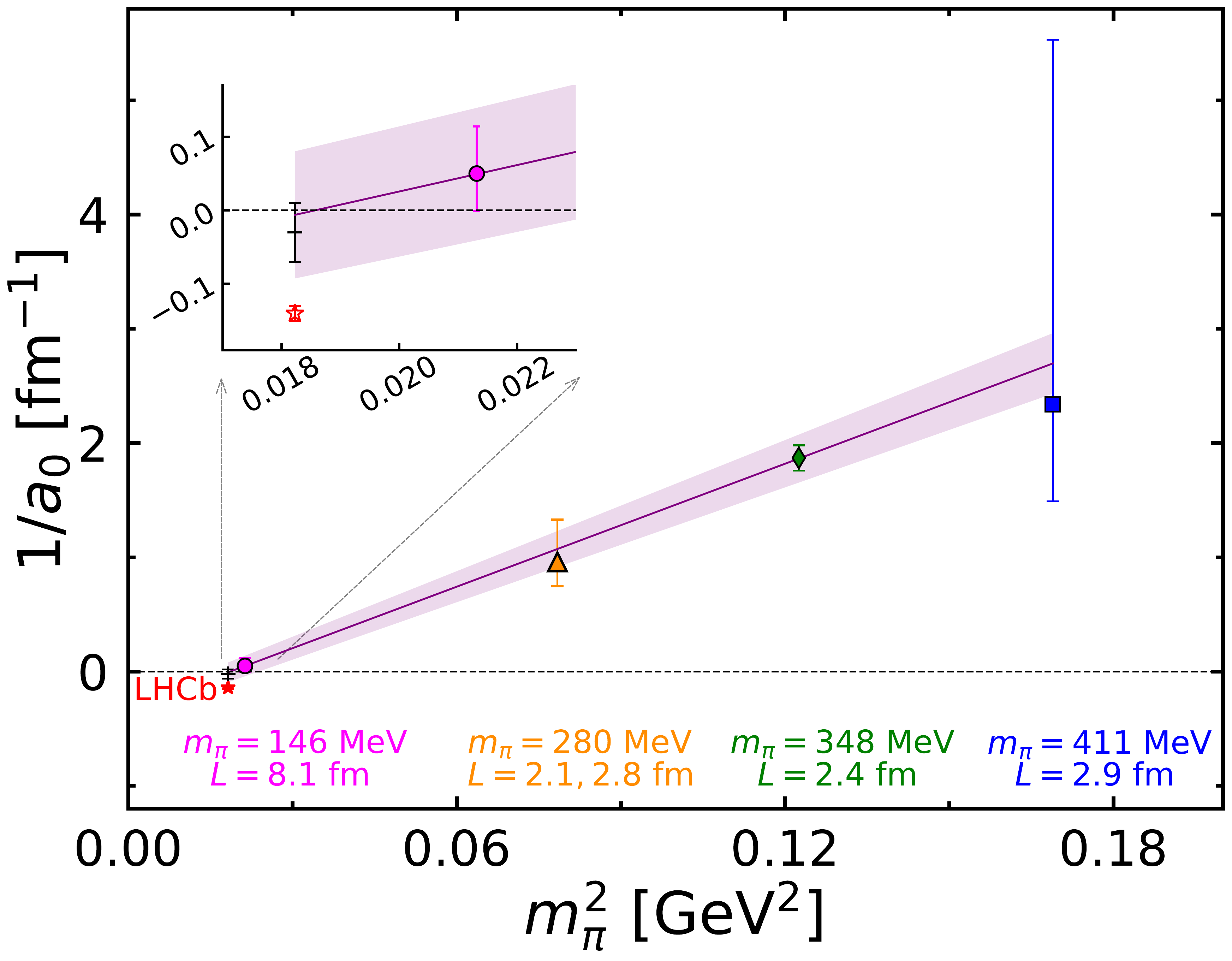}
  \caption{
 The chiral extrapolation of lattice data for $1/a_0$ based on fit function in Eq.~(\ref{eq-mpi-fit}).
The black plus symbol shows the value from the potential modified to $m_\pi=135$ MeV.}
\label{Supp-Fig2}
\end{figure}

Shown in Fig.~\ref{Supp-Fig2} is a chiral extrapolation of lattice data for $1/a_0$ based on following fit function,
\begin{eqnarray}\label{eq-mpi-fit}
 1/a_0(m_\pi)=c+d m_\pi^2.
\end{eqnarray}
Fit parameters $c$ and $d$ together with corresponding $1/a_0$ for $m_\pi=135.0$ MeV are shown in Table~\ref{tab-mpi-fit}.
Result for $1/a_0$ from fit function agrees with the value of $1/a_0 = -0.03(4)~\text{fm}^{-1}$ from the potential modified to the same pion mass according to the TPE, which indicates that the modification of the potential is reasonable.

\end{document}